\documentclass[%
 reprint,
 superscriptaddress,
 amsmath,amssymb,
 aps,
]{revtex4-1}

\usepackage{graphicx}
\usepackage{dcolumn}
\usepackage{bm}
\usepackage{textcomp}
\usepackage{gensymb}

\begin{document}


\title{Complex magnetic order in YbMn$_2$Sb$_2$ single crystals observed by $\mu$SR}

\author{J.~Munevar}
\email{Corresponding author: julian.munevar@ufabc.edu.br}
\affiliation{CCNH, Universidade Federal do ABC (UFABC), Santo Andr\'e, SP 09210-580, Brazil}
\affiliation{Laboratory for Muon Spin Spectroscopy, Paul Scherrer Institut, 5232 Villigen PSI, Switzerland}
\author{F.~R.~Arantes}%
\author{L.~Mendon\c{c}a-Ferreira}
\author{M.~A.~Avila}
\affiliation{CCNH, Universidade Federal do ABC (UFABC), Santo Andr\'e, SP 09210-580, Brazil}
\author{R.~A.~Ribeiro}
\affiliation{CCNH, Universidade Federal do ABC (UFABC), Santo Andr\'e, SP 09210-580, Brazil}
\affiliation{Department of Physics and Astronomy, Iowa State University, Ames, Iowa 50011, USA}



\date{\today}

\begin{abstract}

The crystal growth and the structural, transport and magnetic properties of the magnetically frustrated YbMn$_2$Sb$_2$ single crystals are reported.  The crystals show a trigonal symmetry (space group $P\bar{3}m1$), where corrugated honeycomb layers of MnSb are separated by Yb atoms.  No structural phase transition was observed down to 22~K.  The resistivity measurements show a predominantly insulating behavior.  The combined resistivity, dc susceptibility and heat capacity measurements confirm successive transitions at 230~K, 116~K and 27~K, being the transition at $T_N$=116~K due to the Mn$^{+2}$ lattice antiferromagnetic ordering.  Muon spin rotation experiments ($\mu$SR) reveal a more complicated scenario, with temperature dependence of the magnetic volume fraction reflecting short range and long range magnetic order, and a strongly disordered magnetic ground state.  The role of spin-lattice coupling and its relationship with exchange interactions between Mn moments are discussed as possible cause of the complex magnetic behavior observed.

\end{abstract}

\pacs{Valid PACS appear here}
\maketitle


\section{Introduction}
 
\begin{figure}[!htb]
\begin{center}
\includegraphics[width=8.5cm]{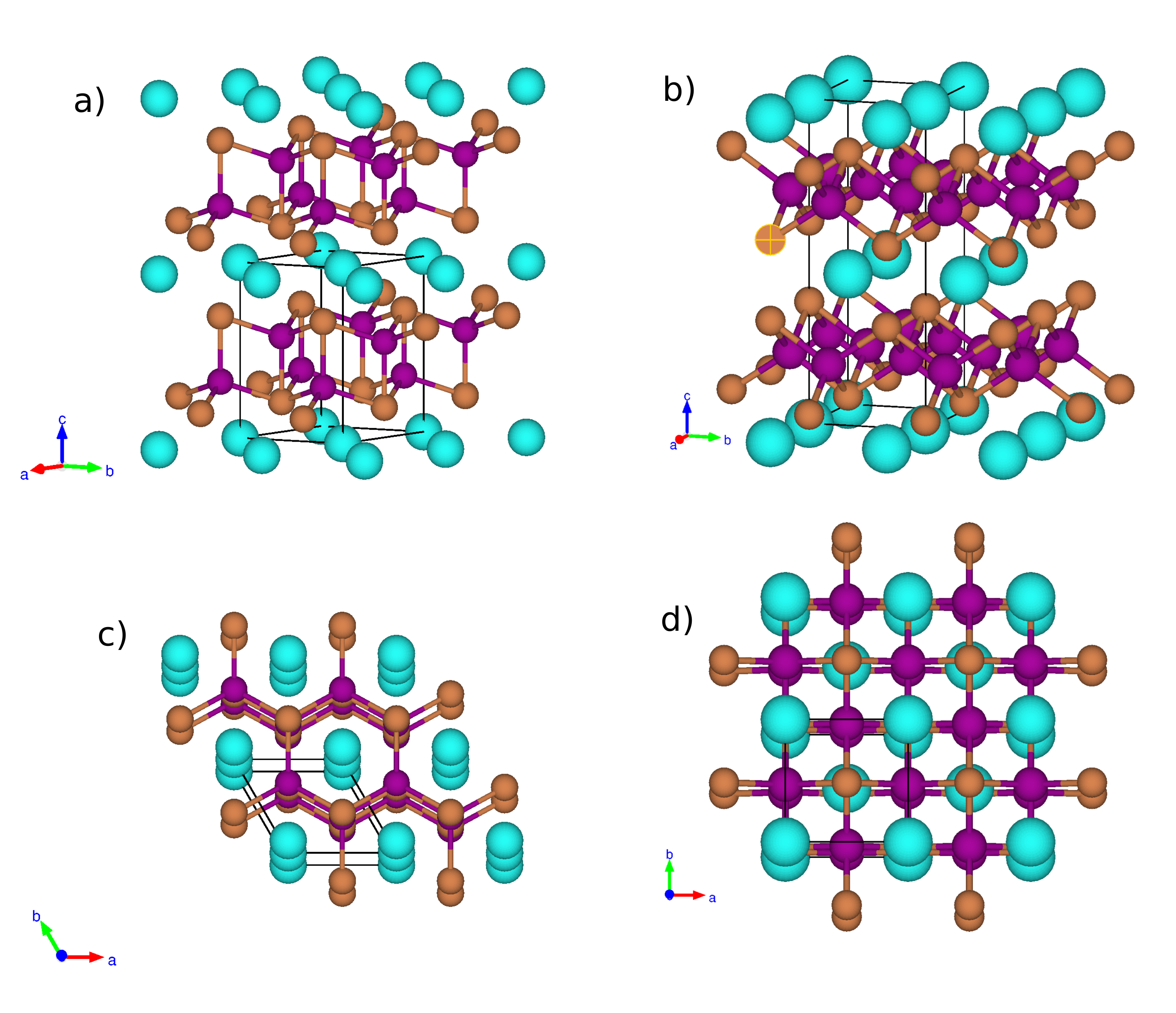}
\caption{Crystal structure for (a) trigonal YbMn$_2$Sb$_2$ and (b) tetragonal YbMn$_2$Si$_2$ \cite{rossi1978structure}.  Views of the crystal lattice perpendicular to the $ab$ plane for each structure are shown in (c) and (d).  Yb atoms are represented in light blue, Mn atoms in purple and Sb/Si atoms in brown.  The lines between the Mn and Si/Sb atoms represent the corresponding bonds.}
\label{Fig0}
\end{center}
\end{figure}

The family of compounds with general formula AM$_2$X$_2$, where A is a divalent cation, M is a transition metal and X are pnictides or chalcogenides, has recently gained interest due to their complex magnetic properties \cite{szytula1981magnetic, simonson2012magnetic, anand2016metallic}, and to their potential to yield new high-temperature superconductors \cite{sasmal2008superconducting, torikachvili2008pressure, anand2015mu} or thermoelectric materials \cite{gascoin2005zintl, zhang2008new}.  This family of compounds can show either the tetragonal ThCr$_2$Si$_2$ structure as in the Fe-based superconductors, or the trigonal CaAl$_2$Si$_2$ structure, where the MX layer presents a corrugated honeycomb lattice, known as a lattice where frustrated magnetism is commonly observed.  Fig. \ref{Fig0} illustrates the difference in the MnSb layer due to crystal lattice symmetry \cite{rossi1978structure}.  Frustrated magnetism has been thoroughly investigated during the last decades \cite{balents2010}, since it can come from geometric constraints in the crystal lattice or from complex behavior of the competing exchange interactions between moments.  In the honeycomb lattice, the magnetic moments lie in the vertices of a hexagon and, in this arrangement, magnetically frustrated interactions arise from competition of the exchange interactions between nearest neighbours (n.n) $J_1$, next-nearest neighbours (n.n.n) $J_2$, the third next-nearest neighbours $J_3$, and even exchange between moments in different layers along the $c$ axis $J_c$.

In the AM$_2$X$_2$ compounds where the transition metal is Mn, the magnetic properties and their corresponding critical temperatures strongly depend on the crystalline structure, since they can crystallize either in the tetragonal body-centered structure ThCr$_2$Si$_2$ or in the trigonal CaAl$_2$Si$_2$ structure  \cite{zheng1988complementary}.  In the first case, the magnetic order of Mn is only destroyed at several hundred Kelvin, as in YbMn$_2$Si$_2$ and BaMn$_2$Bi$_2$, with N\'eel temperatures ($T_N$) of 526~K and 390~K, respectively \cite{hofmann2001magnetic, calder2014magnetic}.  In the latter case, compounds with trigonal structure exhibit antiferromagnetic (AFM) order of Mn moments at distinctly lower temperatures ($<160$~K), and in some cases a weak ferromagnetism above this transition temperature has been reported \cite{anand2016metallic, simonson2012magnetic}.  In both structures the [M$_2$X$_2$]$^{-2}$ lattices are separated by layers of A$^{+2}$ ions; however, the Mn atoms are arranged in a corrugated structure when the compounds display a trigonal symmetry, whereas for the tetragonal symmetry there is a two-dimensional planar lattice (see Fig. \ref{Fig0}).

One of the compounds that has recently attracted attention, CaMn$_2$Sb$_2$, crystallizes in the trigonal structure with $T_N$ between 85-88~K and with Mn moments aligned antiferromagnetically in the \emph{ab} plane \cite{ratcliff2009magnetic, bridges2009magnetic}.  Magnetization measurements suggest weak ferromagnetism above $T_N$ coming from magnetic polarons formed at crystal defects \cite{simonson2012magnetic}.  It is argued that the low $T_N$ for this compound is caused by a frustrated antiferromagnetic honeycomb lattice, where the magnetic order could be one of three possible magnetic phases, namely, a N\'eel phase and two spiral phases \cite{mcnally2015camn}, i.e., the magnetically ordered state is nearly degenerate.  This hypothesis, however, is disputed in a recent work, which assigns the suppression of the ordering temperature to a low dimensionality of the Mn network that leads to a frustrated bond in one of the directions in the \emph{ab} plane \cite{anand2016metallic}.

The compound YbMn$_2$Sb$_2$ studied in this work crystallizes in the trigonal structure mentioned above for CaMn$_2$Sb$_2$.  For this compound, a recent study in polycrystalline samples claims a ferromagnetic (FM) phase at room temperature, with a Curie temperature ($T_C$) of 338~K and a magnetic moment of about 3.5 $\mu_B$/Mn \cite{nirmala2005magnetism}, followed by an antiferromagnetic ordering of Mn below 120~K, with no evidence of magnetic moment for Yb  down to 4~K \cite{morozkin2006synthesis}.  Resistivity, thermal conductivity and Seebeck effect measurements confirm the antiferromagnetic transition observed in the magnetization measurements, showing also an intriguing metallic behavior and a high thermal conductivity due to lattice vibrations \cite{nikiforov2014anomalies}.

These strikingly different results can be settled by the use of single crystalline specimens of YbMn$_{2}$Sb$_{2}$.  The growth of single crystals of intermetallic YbMn$_{2}$Sb$_{2}$ also enables the study of the magnetic properties for different crystal orientations, and for this reason we were motivated to pursue the growth of intermetallic single crystals by the flux method \cite{ribeiro2012growth}.  To our knowledge, there are no reports on magnetic studies on single crystalline specimens of YbMn$_{2}$Sb$_{2}$ up to the present date. 

Magnetization, resistivity, heat capacity and muon spin rotation ($\mu$SR) experiments on YbMn$_{2}$Sb$_{2}$ single crystals can provide a unique point of view of the realization of the magnetic state of YbMn$_2$Sb$_2$.  In particular, the advantage of using a local probe technique such as $\mu$SR is that it allows access to a fluctuation time window that is not accessible by standard magnetization or neutron scattering measurements.  Its unique sensitivity to small internal fields can detect magnetic order from very weak magnetic moments, allows the observation of disordered magnetic order, the dynamics of fluctuating fields and also the phase separation between magnetic and nonmagnetic phases within the same sample.  The combination of the above mentioned techniques can extend the comprehension of the physical properties of YbMn$_2$Sb$_2$, and will allow a comparison with the previously proposed models for the magnetic order in YbMn$_2$Sb$_2$ and CaMn$_2$Sb$_2$.  We find that the magnetic order evolves from a weak ferromagnetic order to a disordered long range anisotropic antiferromagnetic order as temperature decreases.

\section{Experimental Methods}
High-quality single crystals of YbMn$_2$Sb$_2$ were grown using Sn-flux technique.  Amounts of Yb, Mn, Sb and Sn in the ratio 1:2:2:30 were weighted on a high-precision balance and put into an alumina crucible.  The crucible was inserted into a quartz tube together with quartz wool, and the tube was evacuated and sealed.  The mixture was heated to 1423 K for 4 h and further kept at this temperature for 24 h, and then cooled down to 823 K over 120 h.  The ampoule was taken to a centrifuge to remove the flux and large single crystals of YbMn$_2$Sb$_2$ were extracted. Hexagon-shaped black crystals of approximately 2.5$\times$2.5~mm$^2$ were obtained, as shown in Fig.~\ref{Fig1}.  It is worth noticing that the crystal shape already suggests a trigonal crystal lattice.  To investigate the structural properties, temperature dependent X-ray diffraction patterns from powdered samples were measured in a BRUKER diffractometer with a Mo $K\alpha$ X-ray tube ($\lambda=0.709$ \AA).  The resistivity and specific heat measurements were performed in a Quantum Design PPMS: the DC resistivity between 2~and~300~K was performed on two different crystals of the same batch, and the specific heat with the heat capacity option ranging temperatures from 2 to 250 K.  The dc magnetic properties were measured with a Quantum Design MPMS3 SQUID-VSM magnetometer from 2~to~300~K.  $\mu$SR spectra were obtained using the GPS instrument at the Swiss Muon Source of the Paul Scherrer Institute, Switzerland.  Zero field (ZF), weak transverse field (wTF) and longitudinal field (LF) spectra were obtained for temperatures from 1.6 to 630 K and longitudinal fields up to 150 mT.

\section{Results}

\begin{figure}[!htb]
\begin{center}
\includegraphics[width=8.5cm]{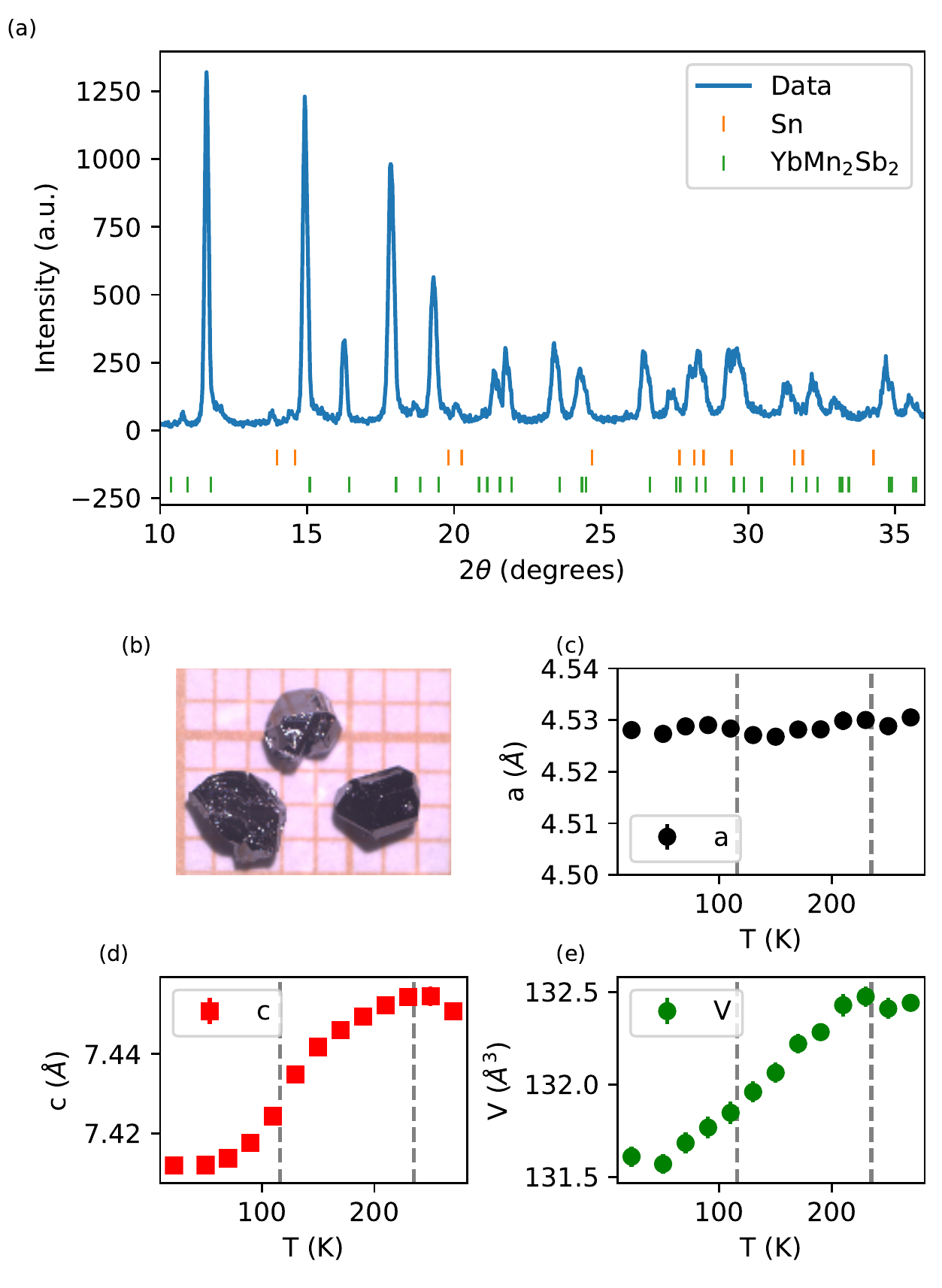}
\caption{(a) X-ray diffraction pattern for YbMn$_2$Sb$_2$ at room temperature, showing the corresponding reflections for the main phase and Sn impurity from the flux.  (b) YbMn$_2$Sb$_2$ single crystals. (c)-(e) Temperature dependence of lattice parameters $a$ and $c$, and crystal lattice volume $V$ obtained from the Rietveld refinement, respectively.}
\label{Fig1}
\end{center}
\end{figure}

The temperature dependent X-ray diffraction patterns confirm a trigonal structure belonging to the space group \emph{P$\bar{3}$m}1 between 22 K and 300 K.  The expected reflections for YbMn$_2$Sb$_2$ and Sn (from the flux) were observed (Fig. \ref{Fig1}(a)): the Rietveld refinement of the pattern at room temperature results in $a=4.5278$~\AA, and $c=7.4481$~\AA, with the presence of about 3\% of Sn.  These lattice parameters are very close to those reported previously \cite{morozkin2006synthesis}, and smaller than the ones reported for CaMn$_2$Sb$_2$ \cite{simonson2012magnetic}, due to the difference in the atomic radius between Ca$^{+2}$ and Yb$^{+3}$.  The Rietveld refinement confirmed that the crystal structure consists of Yb atoms that separate the MnSb layers, where the Mn atoms are connected by Sb atoms and form a corrugated honeycomb lattice.  The Mn-Mn distance obtained at room temperature is d$_{Mn-Mn}$=3.099 \AA, smaller than the d$_{Mn-Mn}$ values for CaMn$_2$Sb$_2$ \cite{simonson2012magnetic}. 

The temperature evolution of the lattice parameters and the unit cell volume were obtained by performing a global fit using the GSASII software \cite{gsasii}, taking all the patterns from 22 to 270 K and the same phases (main phase and Sn impurity).  These results are shown in the Figs.~\ref{Fig1}(c)-(e).  It is clearly seen that the $a$ lattice parameter has a negligible temperature variation, while the $c$ lattice parameter shows a decrease starting below approximately 250 K and becomes more pronounced near 120 K.  The unit cell volume also reflects a similar decrease.  The influence of these changes in the lattice parameters on the magnetic properties and the temperature dependence of d$_{Mn-Mn}$ will be discussed later.

\begin{figure*}[!htb]
\begin{center}
\includegraphics[width=18cm]{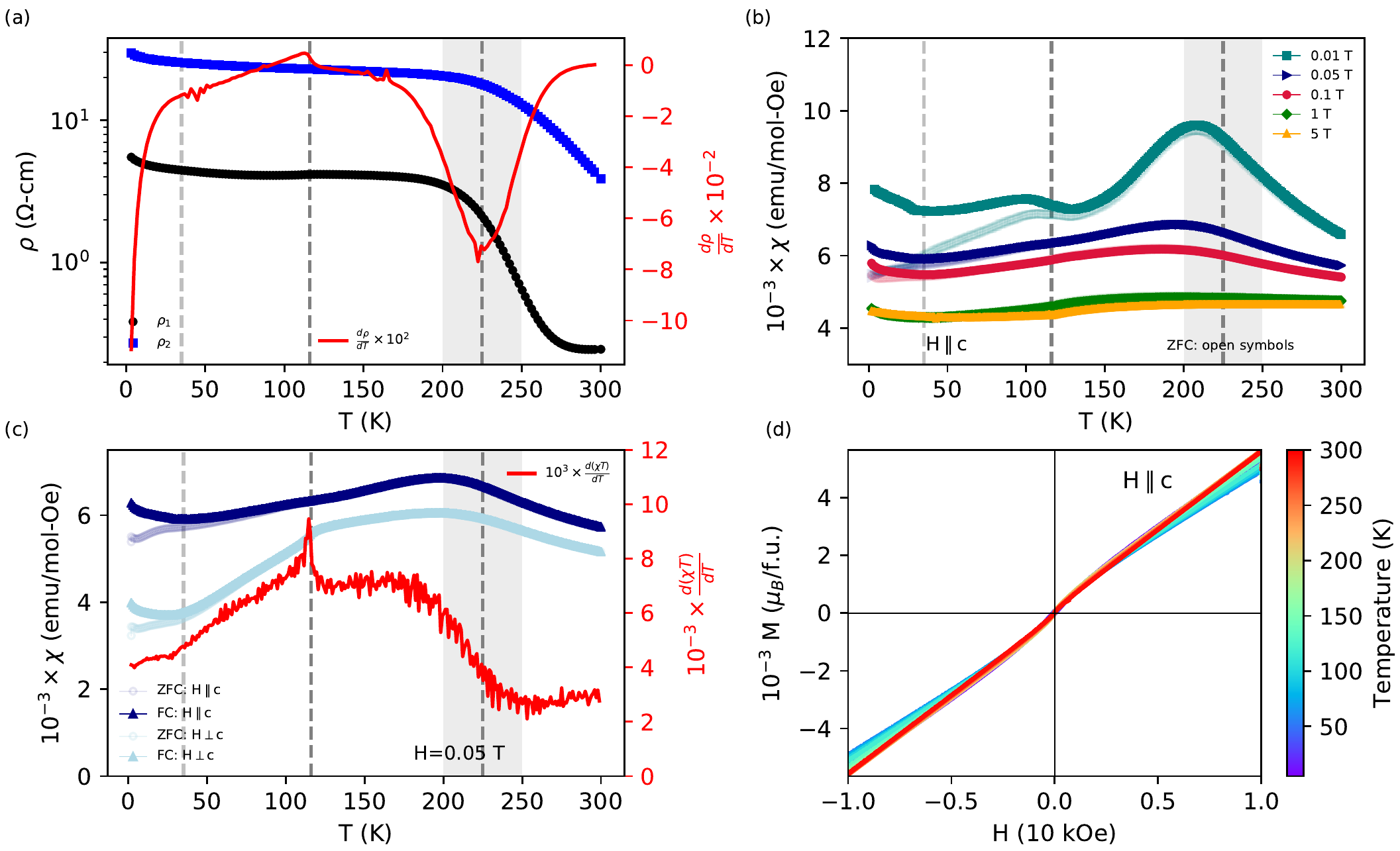}  
\caption{(a) Resistivity measurements performed for two single crystals of YbMn$_2$Sb$_2$ belonging to the same batch.  $\frac{1}{\rho}\frac{d\rho}{dT}$ is also shown in red.  (b) ZFC-FC curves for different magnitudes of applied magnetic field measured with $H\parallel c$. (c) ZFC-FC curves with an applied field of 0.05~T and measured along the axes \emph{a} and emph{c}. Below 116~K the light blue curves ($H\perp c$) show a behavior consistent with an antiferromagnetic ordering. (d) External magnetic field dependence of the magnetization from 5~K up to room temperature with $H\parallel c$.} \label{Fig3}
\end{center}
\end{figure*}

The temperature dependence of the resistivity of YbMn$_{2}$Sb$_{2}$ is shown in the Fig. \ref{Fig3}(a).  Despite showing distinct resistivity values, crystals from the same batch present insulating behavior.  This was previously observed in resistivity measurements on CaMn$_2$Sb$_2$, and it was suggested that crystal defects may be responsible for such behavior.  However, the resistivity in CaMn$_2$Sb$_2$ varies over several decades \citep{simonson2012magnetic}, whereas our measurement does not show such a remarkable increase.  By comparing our results with those reported in ref. \citep{simonson2012magnetic}, we notice that the effect of Yb in the transport properties resembles the hole doping induced by Na substitutions at the Ca site in CaMn$_2$Sb$_2$.  The insulating behavior observed for our two different crystals of the same YbMn$_2$Sb$_2$ batch differs from the results reported in \cite{nikiforov2014anomalies}, where polycrystalline YbMn$_2$Sb$_2$ clearly show metallic behavior.  The effect of impurities on the transport properties may be more significant in polycrystalline specimens \cite{magnavita2016}.  Fig. \ref{Fig3}(a) also shows $\frac{1}{\rho}\frac{d\rho}{dT}$, in particular reflecting a phase transition at 116 K and a minimum at approximately 245 K.



A previous neutron diffraction study on YbMn$_2$Sb$_2$ showed antiferromagnetic order below 120~K, with the Mn magnetic moments forming an angle $\theta\approx$ 64\textdegree{} with respect to the \emph{c}-axis \citep{morozkin2006synthesis}. This magnetic structure is different from the one observed for CaMn$_2$Sb$_2$, where the Mn magnetic moments lie in the \emph{ab} plane \citep{mcnally2015camn,simonson2012magnetic}.  This difference is not obvious, since both crystal lattices are similar.  Also, it was previously reported that YbMn$_2$Sb$_2$ has a clear ferromagnetic order at room temperature \citep{nirmala2005magnetism}.  We will show that the magnetic order in YbMn$_2$Sb$_2$ is much more complex than previously reported.

%

The magnetization measurements for YbMn$_2$Sb$_2$ are shown in Figs.~\ref{Fig3}(b)-(d).  The ZFC-FC curves measured in the direction parallel to the crystallographic \emph{c} axis of the crystal and under different external fields are shown in Fig.~\ref{Fig3}(b).  Three transitions are clearly observed, at approximately 30~K, 116~K and 200~K.  The transition at 116~K is similar to that observed in CaMn$_2$Sb$_2$ and, according to the neutron diffraction studies in CaMn$_2$Sb$_2$ and YbMn$_2$Sb$_2$, antiferromagnetic order is expected. The broad peak around approximately 200~K is very sensitive to the magnitude of the external field, as it shifts to lower temperatures with an increasing external field. The ZFC and FC curves indicate irreversible behavior in the whole temperature range measured, in particular at low external fields.  Finally, all the features for measurements at low fields are suppressed with a large external field, suggesting that the magnetic interactions present are weak.  

Figure \ref{Fig3}(c) shows the ZFC-FC curves with an external field of 0.05~T both parallel and perpendicular to the crystallographic $c$-axis.  There is a continuous increase of $\chi$ from room temperature up to 200~K, followed by a small decrease, and magnetic anisotropy is reflected by the larger $\chi$ values for $H||c$.  The magnetic transition at 116~K is noticed only as a variation in the magnetization in the $ab$ plane, and no remarkable difference between ZFC and FC modes is seen.  The AFM transitions are more clearly seen when $\frac{d\left(\chi_e T\right)}{dT}$ is plotted (see Fig. \ref{Fig3}(c), red line), where $\chi_e=\frac{1}{3}\chi_{\parallel}+\frac{2}{3}\chi_{\perp}$.  Below 30~K, a further transition is seen as an irreversible behavior between the ZFC and FC measurements, possibly caused by a rearrangement of Mn magnetic moments.  


Figure \ref{Fig3}(d) shows $M\times H$ curves with the external field parallel to the crystallographic $c$ axis at different temperatures.  This figure shows an interesting behavior: a decrease of the magnetization as the temperature decreases, possibly caused by the crossover between the weak ferromagnetic phase and the AFM phase.  The Mn atoms reach a small magnetic moment of approximately $10^{-2}\mu_B$/f.u. without the observation of metamagnetic transitions (curves measured at 5~K and 100~K), and with a small but noticeable difference between different crystal orientations (not shown), indicative of a small magnetic anisotropy related to the magnetic order.  

\begin{figure}[!htb]
\begin{center}
\includegraphics[width=9cm]{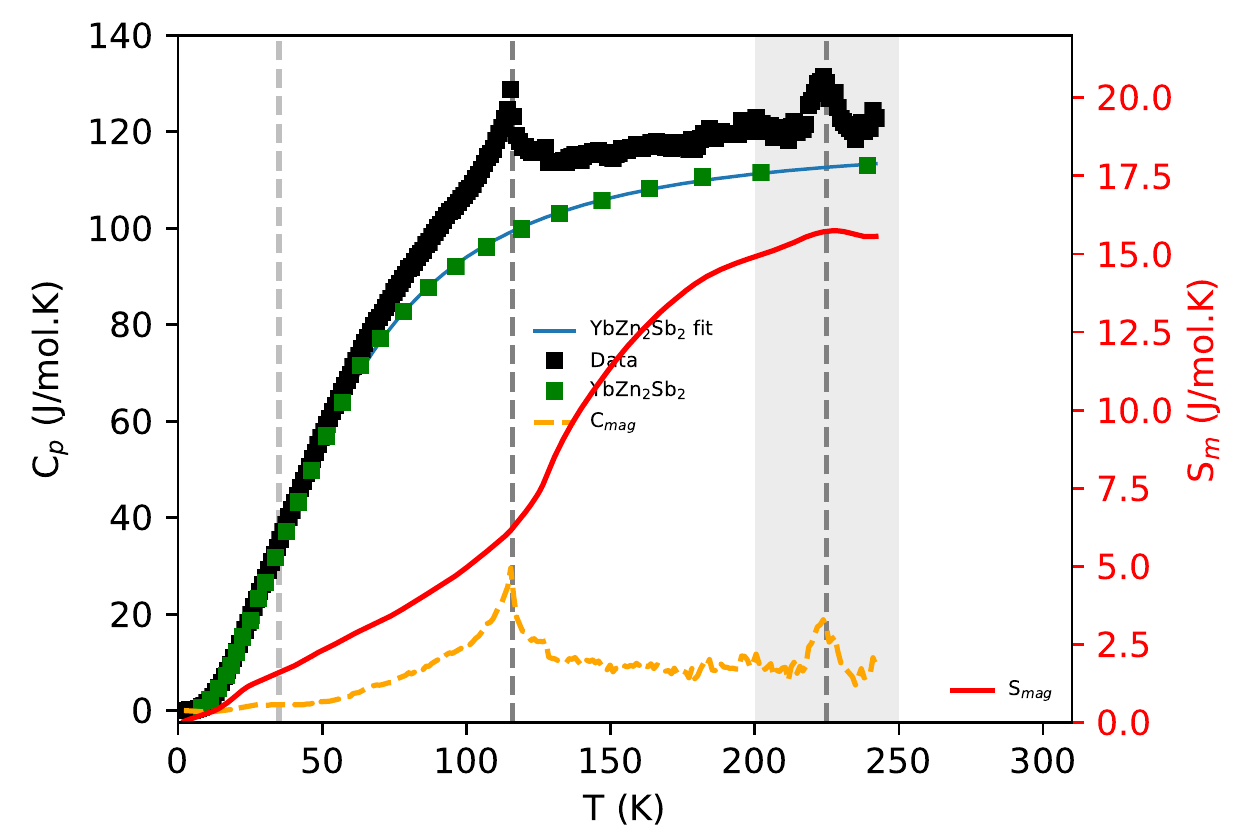}
\caption{Specific heat as a function of the temperature for YbMn$_2$Sb$_2$. The peak at 116~K indicates the antiferromagnetic transition.  The magnetic heat capacity and the respective magnetic entropy are also shown, with a line showing the approximate value of $S_{mag}$ for $S=\frac{5}{2}$.} \label{Fig_Cp}
\end{center}
\end{figure}

The temperature dependent specific heat measurement (Fig. \ref{Fig_Cp}) shows two magnetic transitions at approximately 235~K and 116~K, related to the weak FM and AFM transitions observed from magnetization measurements, respectively.  Interestingly, no evidence of further magnetic transitions close to 30~K is observed.  The low temperature fit of the specific heat ($C_P/T$ versus $T^2$) yielded a Debye temperature of $\theta_D=106$ K, which is significantly lower than that reported for CaMn$_2$Sb$_2$ \citep{simonson2012magnetic}.  In order to extract the magnetic contribution to the sample specific heat, we took the specific heat data of the non-magnetic equivalent of YbMn$_2$Sb$_2$, i.e. YbZn$_2$Sb$_2$ \cite{may2012ybzn2sb2}, and subtracted it from the data obtained for YbMn$_2$Sb$_2$.  The specific heat curve for YbZn$_2$Sb$_2$ was modelled using a sum of two contributions, namely, one following the Debye model and the other following the Einstein model: 

\begin{align*}
C_p(T)&=f\times9Nk\left(\frac{T}{\theta_D}\right)^3\int_0^{\frac{\theta_D}{T}}dxe^xx^4/(e^x-1)^2+\\&+(1-f)\times3Nk\left(\frac{\epsilon}{kT}\right)^2\frac{e^{(\frac{\epsilon}{kT})}}{\left(e^{(\frac{\epsilon}{kT})}-1\right)^2},
\label{eq_debye}
\end{align*}

where $N$ is the Avogadro constant, $k$ is the Boltzmann constant, $\theta_D$ is the Debye temperature, $T_E$ is the Einstein temperature, $\epsilon=kT_E$ and $T$ is the temperature.  The corresponding Debye and Einstein temperatures are $\theta_D^{Zn}=131(1)$~K and $\theta_E^{Zn}=235(3)$~K.

We separated the lattice and magnetic contributions to the heat capacity, and integrated the magnetic heat capacity ($C_{mag}$) in order to obtain the magnetic entropy ($S_{mag}$, Fig.~\ref{Fig3_Cp}, red curve).  Our analysis yields $S_{mag}=$14.46 J/mol-K, which corresponds to $S=2.35\simeq\frac{5}{2}$ ($S_{mag}=R\ln{\left(2S+1\right)}$), meaning that the Mn moments are in the +2 high-spin state.

The $\mu$SR experiments were conducted on a single crystal with its $c$ axis parallel to the muon beam (i.e. muon momentum), and the muon spin rotated 40 degrees relative to the muon beam.  Muon decays were detected by four detectors labeled as up (U), down (D), backward (B) and forward (F).  Using this setup, the magnetic response related to the crystallographic $c$ axis (UD), as well as to the $ab$ plane (BF) was observed, since each set of detectors reflect the projection of the muon spin polarization influenced by the internal field in the corresponding direction.


\begin{figure}[!htb]
\begin{center}
\includegraphics[width=8.5cm]{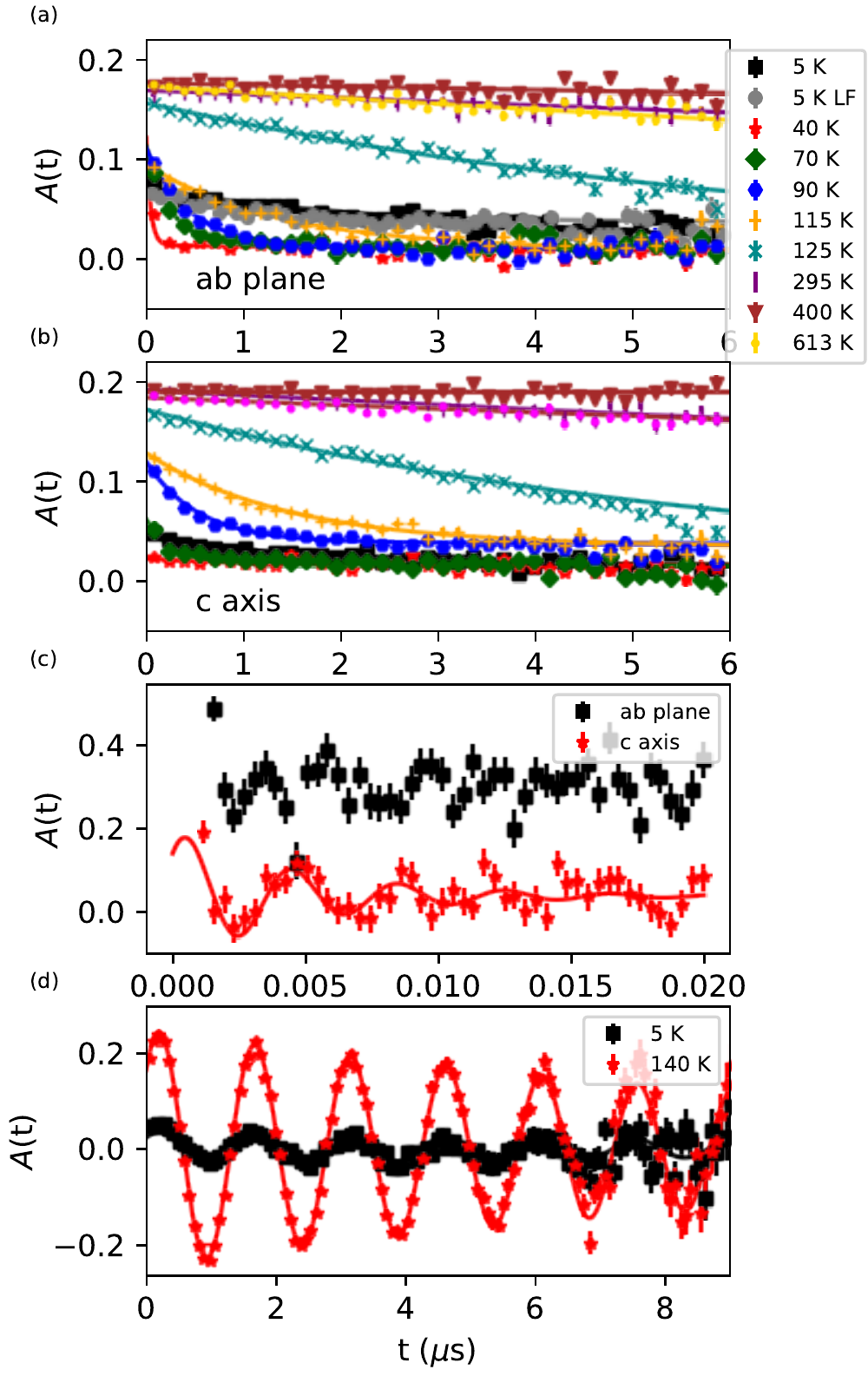}
\caption{ZF-$\mu$SR spectra obtained for YbMn$_2$Sb$_2$ single crystal between 5 and 613 K, for (a) the BF detectors (muon spin projection in the crystal $ab$ plane) and (b) the UD detectors (muon spin projection in the crystal $c$ axis) detectors. (c) ZF-$\mu$SR spectra at 5~K at early times ($<20$ ns).  (d)~5~mT weak TF-$\mu$SR spectra at 5 and 140 K.} \label{Fig6}
\end{center}
\end{figure} 

The general fitting function for an asymmetry spectrum of a single crystalline specimen is strongly dependent on the initial muon spin and the internal field directions.  In a simplified case, where the internal fields are assumed to be static and magnetic order is long range, it is given by 
\begin{equation}
A(t)=A_0\left(e^{-\lambda_L t}\cos^2{\theta}+e^{-\frac{\sigma_T t}{2}}\sin^2{\theta}\cos{\left(\omega t\right)}\right) ,\label{eq1}
\end{equation}

where $A_0$ is the initial asymmetry, $\lambda_L$ is the relaxation rate for the longitudinal component of the asymmetry, $\sigma_T$ is the Gaussian relaxation rate for the transverse component of the asymmetry, $\theta$ is the angle between the internal field and the muon direction, and $\omega=\gamma_{\mu}B_{\mu}$ is the muon spin angular frequency.  Equation \ref{eq1} was used to fit the spectrum in Fig. \ref{Fig6}(c).  A different fitting function was chosen for the spectra in Figs. \ref{Fig6}(a)-(b) due to the absence of muon spin precession, the evidently exponential relaxation seen, and assuming a quasistatic regime:


\begin{equation}
A_0 P(t)=A_s e^{-\lambda_s t} + A_F e^{-\lambda_F t} +\text{BG}. \label{eq2}
\end{equation}

$A_{0}$ is the initial asymmetry of the muon, $A_s$ is the asymmetry corresponding to a component reflecting slow muon spin relaxation $\lambda_{s}$, $A_F$ is the asymmetry corresponding to a component of the spectrum reflecting a fast relaxation rate $\lambda_F$, and BG is a temperature independent background contribution to the total polarization.  

The fitting function for the weak TF-$\mu$SR spectra (Fig. \ref{Fig6}(d)) is:


\begin{equation}
P(t)=P_{TF}e^{-\lambda_{TF} t}\cos{\left(\gamma_{\mu}B_{TF}t\right)}, \label{eq3}
\end{equation}

where $P_{TF}$ is the muon polarization under the external weak transverse field, $\lambda_{TF}$ is the corresponding relaxation rate and $B_{TF}$ is the total field sensed at the muon site.  All the spectra were fitted within the 0.2-10~$\mu$s time range.  Global fits were performed using the \textsc{Musrfit} software \cite{musrfit}, keeping $\lambda_F$ and BG constant over $5-117$~K.

Figures \ref{Fig6}(a)-(b) show selected ZF-$\mu$SR spectra for temperatures from 5 K to 613 K.  An initial decrease of asymmetry accompanied by a temperature-dependent exponential damping of the asymmetry is observed for the full temperature range.  An asymmetry decrease of approximately 12 \% is presumed to occur between 140 and 295 K in both orientations.  A further decrease of the asymmetry is observed below approximately 116 K, reflecting the robustness of the long range antiferromagnetic transition previously observed.  Figure \ref{Fig6}(c) shows early time ZF-$\mu$SR spectra for both orientations ($\sim$20 ns) at 5~K.  A clear muon spin precession was seen for the muon spin projection on the $c$ axis only, where it is also observed that for both spectra the asymmetry is strongly damped at very early times ($\sim$ 10 ns).  This could explain the absence of muon spin precession for higher temperatures.

The precessing spectrum in Fig. \ref{Fig6}(c) was fitted using Eq. \ref{eq1}, setting $\lambda_L$ equal to zero.  The internal field obtained is $B_{int}=1.84(6)$~T, $\sin^2{\theta}=0.81(3)$ and $\sigma_F=194(50)$~$\mu s^{-1}$.  The muons stopping in the crystal are sensing a very strong and broadened field, and it can be inferred that the internal field direction with respect to the muon spin polarization (or the crystal $c$ axis) is $\theta=64(1)$ degrees, which is consistent with the previous neutron study \citep{morozkin2006synthesis}.

\begin{figure}[!htb]
\begin{center}
\includegraphics[width=8.7cm]{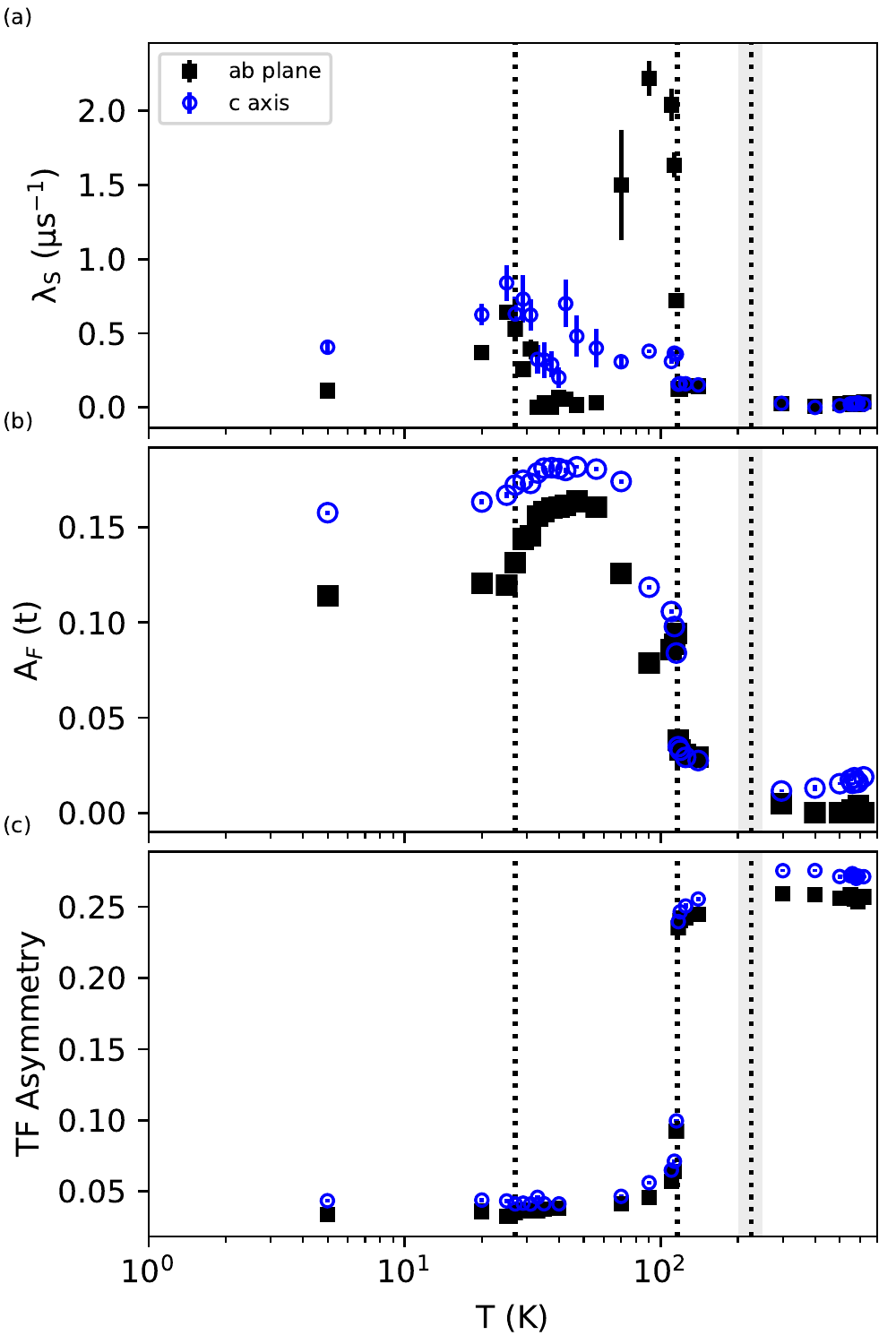}
\caption{(a) Muon spin relaxation rate, (b) average internal field angle and (c) TF asymmetry for YbMn$_2$Sb$_2$ single crystal.  The dashed lines indicate the temperatures where a magnetic transition is observed/expected.} \label{Fig7}
\end{center}
\end{figure}

The relaxation rates of the slowly relaxing component $\lambda_{s}$ obtained from ZF-$\mu$SR, the zero field asymmetries of the fast component $A_F$, and the magnetic volume fraction obtained from weak TF-$\mu$SR measurements are shown in Fig. \ref{Fig7}(a)-(c).  The behavior of $\lambda_s$ (Fig. \ref{Fig7}(a)) shows a minor increase as it approaches to $T_N$, as is expected from the slowing down of the moment fluctuations.   Below 116~K, $\lambda_s$ shows different behavior for each projection of the muon spin explored, as follows: in the $ab$ plane is seen a considerable increase of $\lambda_s$ is seen, reaching a maximum at approximately 80 K and decreasing, whereas for the $c$ axis a modest increase is seen at $T_N$, followed by a further increase below approximately 70 K.  Furthermore, an additional increase in $\lambda_s$ is observed at lower temperatures with peak at approximately 27~K. 

Figure \ref{Fig7}(b) shows the ZF asymmetry obtained for the fast relaxing component in the spectra $A_F$, for each muon spin projection.  We see that prior to the onset of the AFM transition at $T_N$, there is an asymmetry increase of about 10 \%.  This asymmetry increase may be associated to regions of the crystal developing short range magnetic order and a very large muon relaxation rate $\lambda_F$.  At $T_N$ we clearly see a jump in $A_F$ reflecting the magnetic transition, and below $T_N$, we observe a further increase of $A_F$ reaching a maximum at approximately 40 K and stabilizing below 27~K.  In order to extract the nonmagnetic volume fraction of the sample, we analysed the weak TF-$\mu$SR spectra (Fig. \ref{Fig6}(d)), and the TF asymmetry extracted is shown in Fig. \ref{Fig7}(c).  This TF asymmetry is proportional to the nonmagnetic volume fraction of the crystal, and it starts to decrease already at high temperatures ($\sim$300~K), being followed by a sharp decrease at 116 K and reaching a minimum value at lower temperatures.  Figure \ref{Fig7}(c) implies that below $T_N$ the system undergoes a magnetic phase transition, which is consistent with our previous results.  


In order to verify whether the internal fields are static or dynamic, LF-$\mu$SR experiments were performed at 5 K for fields as large as 0.15~T (see Fig. \ref{Fig6}).  Our experiments show that the external field has a negligible effect on the muon spin depolarization.  This may be caused because to decouple such strongly interacting spins it would be necessary to apply large fields.  For instance, if we have $\sigma_F\simeq 200$~$\mu s^{-1}$, the LF field necessary to observe asymmetry decoupling would be around $\frac{\sigma_F}{\gamma_{\mu}}\simeq 20$~kG.

\section{Discussion}

We now discuss the different features observed in our analysis, namely, the transitions at $\sim$250~K, at $T_N=116$~K and at $\sim$27~K.  The magnetization measurements in Figs.\ref{Fig3}(b)-(d) at high temperatures show irreversible and anisotropic behavior, suggesting that magnetic order is already present at these temperatures.  Furthermore, the Fig.~\ref{Fig3_Cp} clearly shows a peak in the specific heat indicating a phase transition at around 225~K.  A variation in the muon spin polarization between 140 and 400~K is also seen (Figs. \ref{Fig7}(b) and \ref{Fig7}(c)), reflecting the appearance of a small magnetic volume fraction ($\sim$10~\%) within this temperature range.  

The influence of crystal defects that favor the onset of these observed short range correlations, possibly leading to the previously reported polaron-mediated magnetic interactions can be the reason to observe such behavior \cite{simonson2012magnetic}.  Any magnetically ordered state for $T>$120~K was observed by the neutron diffraction experiments reported in Ref. \cite{morozkin2006synthesis}.  $\mu$SR is a local probe technique that is more suitable to detect magnetic volume fractions, even for small fields or short range order, and that can be the reason for the absence of evidence of magnetic correlations from neutron diffraction.  

The resistivity measurements (see Fig. \ref{Fig3}(a)) evidence an anomalous insulating behavior, with a very broad increase around 225~K.  This behavior may be signature of an increasing conduction electron scattering in the crystal.  The observed behavior of the transport and magnetic properties has been previously assigned to scattering processes induced by the moments present in the sample at these temperature ranges \cite{nikiforov2014anomalies,simonson2012magnetic,bridges2009magnetic}, probably related to paramagnon states at point defects in the crystal lattice above $T_N$ that are predicted in the theory of classical Heisenberg spins interacting in the honeycomb lattice \cite{rastelli1979theory}.  This would explain the observed variation of the magnetization and resistivity between samples up to one order of magnitude, as it was also noted in ref. \cite{simonson2012magnetic}.  This phenomenon could result from the distribution of defects in the crystalline structure varying from one sample to another, and thus affecting the formation of magnetic polarons and in consequence the magnetic response in these temperature ranges. 

It was previously known from neutron diffraction results that the Mn moments align antiferromagnetically below 120 K \cite{morozkin2006synthesis}.  We observe the onset of a long range magnetic order at 116 K, manifested as a phase transition in the resistivity, magnetization and specific heat measurements (Fig. \ref{Fig3} and Fig. \ref{Fig_Cp}), and as a clear increase in $\lambda_s$ and the magnetic volume fraction in the $\mu$SR results (Fig.~\ref{Fig7}).  The magnetization measurements show that the AFM order below 116~K is sensitive to the external magnetic field.  The specific heat confirms the magnetic transition by observing a clear peak at this temperature, the spin state of Mn moments is found to be $S=\frac{5}{2}$.  A higher $T_N$ compared to CaMn$_2$Sb$_2$ ($T_N=85$~K \cite{bridges2009magnetic}) seems plausible, since the distance between Mn atoms is smaller for YbMn$_2$Sb$_2$, in favor of a stronger exchange interaction between Mn moments.

The $\mu$SR spectra do not show any indication of muon spin precession except at 5~K, nor an internal field decoupling due to the longitudinal field at 5~K (see Fig. \ref{Fig6}).  This points to a scenario where static order is present, but strongly disordered.  It is not a spin glass transition because the phase transition also manifests in the specific heat measurement (Fig. \ref{Fig3_Cp}). Anisotropy is also observed from $\mu$SR and magnetization measurements.  Specifically for the $\mu$SR results (see Fig. \ref{Fig7}), a larger $\lambda_s$ is observed for the muon spin projection on the $ab$ plane, suggesting that larger internal field fluctuations occur, which could be taken as an indication of stronger competition within magnetic moments between the $ab$ plane.

A competition between magnetic ground states may be plausible, as follows: the $J_1/J_2$-$J_3$ phase diagram \cite{mcnally2015camn,simonson2012magnetic,rastelli1979theory,mazin2013theory} suggests a plethora of possible magnetic ground states, and the ratio between the corresponding exchange interactions $J_i$ will determine the nature of the magnetic ground state.  A strongly disordered ground state could appear if the system is on the vicinity of several magnetic ground states, as it was demonstrated for CaMn$_2$Sb$_2$ single crystals and its proximity to a magnetic tricritical point \cite{simonson2012magnetic,mcnally2015camn}.  According to our results, this may be the scenario of the magnetic ground state for YbMn$_2$Sb$_2$.

\begin{figure}[!htb]
\begin{center}
\includegraphics[width=8.5cm]{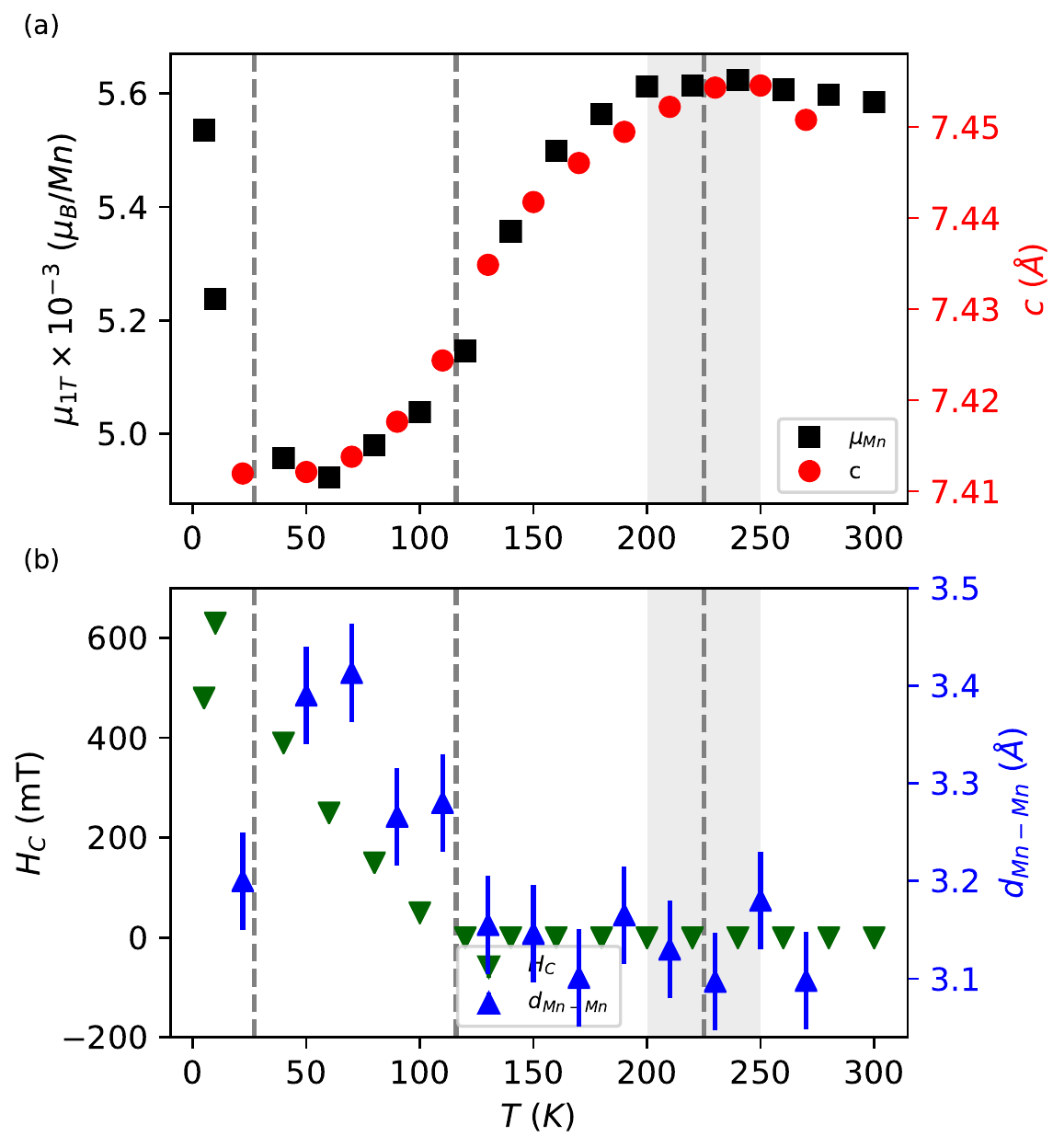}
\caption{(a) Saturation moment at 1 T and $c$ lattice parameters obtained for YbMn$_2$Sb$_2$.  (b) Coercive field and distance between nearest Mn atoms.} \label{Fig8}
\end{center}
\end{figure}

We also found a temperature dependent correlation between the magnetic order and the crystal lattice volume, as can be seen in Fig. \ref{Fig8}(a), where the unit cell parameter $c$ (Fig. \ref{Fig1}(c)) follows a similar behavior compared to the saturation magnetic moment $H_C$, extracted from the hysteresis loops presented in Fig. \ref{Fig3}(d).  There is only a negligible variation of the above mentioned parameters down to 225 K, followed by a decrease reaching a minimum value at around 50 K.  This spin-lattice coupling was suggested by a previous report \citep{nikiforov2014anomalies}, where thermal conductivity measurements indicated spin-lattice interactions accounting for a maximum near 30 K.  This spin-lattice coupling may affect the bulk properties, possibly depending on the crystal size and quality, explaining why the results can vary between samples.  We also see the distance between Mn atoms and the coercive field, plotted in Fig. \ref{Fig8}(b), increasing below approximately 120 K.  

The results presented in Fig. \ref{Fig8} suggest that the compression of the crystal lattice and the subsequent reduction in $d_{Mn-Mn}$ are directly related to the change in the magnetic properties, namely, the irreversible behavior reflected by the increase in $H_C$.  One possible explanation for this behavior may be a spin Jahn-Teller effect that has been discussed previously in magnetically frustrated systems \cite{olariu2006jahnteller,sushkov2005jahnteller,wang2008jahnteller,valdes2008jahnteller}, where a non-Kramers degeneracy may lead to a Jahn-Teller distortion.  A deeper insight is needed in order to clarify the mechanisms involved.

Finally, below 30 K, there is a noticeable increase of the magnetization irreversibility, a maximum in the muon spin relaxation rate $\lambda_s$ and a constant value for $A_F$.  This behavior might be caused by a transition from a N\'eel state to another magnetic state.  According to the change in the lattice parameters and the distances between Mn atoms $d_{Mn-Mn}$, this could be explained by assuming that a temperature dependent reduction of the unit cell dimensions would lead to an increase in $J_c$ and a reduction in $J_1$, $J_2$ and $J_3$, therefore changing the ratios $J_c/J_1$, $J_3/J_1$ and $J_2/J_1$ in favor of another ordered state, as represented in the phase diagram \cite{mcnally2015camn}.  This change would represent no change of entropy in the system and thus no change in the sample specific heat.  However, since the present experiments do not allow extraction of individual exchange constants, it is necessary to seek a more suitable technique to confirm this hypothesis.

The behavior of $\lambda_s$ could be understood within the assumption of a quasistatic scenario, where $\lambda_s\simeq\frac{2\nu}{3}$, and $\nu$ representing the fluctuation rate of the muon spin \cite{dunsiger1996muon,rovers2002prb,takeshita2007jpcs}.  The fluctuation time $\tau=\frac{1}{\nu}$ is around $10^{-6}$~s, close to the resolution limit of $\mu$SR, but slow enough to observe static magnetic order by neutron diffraction.  However, looking Fig. \ref{Fig7}(a) in detail, we see that anisotropy is reflected as a slower fluctuation time $\tau$ close to $T_N$ in the $ab$ plane, and a similar $\tau$ at 27~K.  One possibility would be the crossover from a magnetic ground state where magnetic interactions occur in layers, to a 3D magnetic ground state.  This hypothesis may be supported by the spin-lattice coupling we observe in Fig. \ref{Fig8}, but further experiments are needed to confirm this.

All these observations suggest that the magnetic order observed in YbMn$_2$Sb$_2$ arises from competing exchange interactions between Mn moments controlled by a spin-lattice coupling.  The presence of slowly fluctuating local fields also suggest that the magnetic ground state is strongly disordered, in accordance with previous reports on other materials showing similar behavior \cite{dunsiger1996muon,rovers2002prb,takeshita2007jpcs}.  The observation of magnetic fluctuations at high temperatures, reflected by irreversible magnetization curves at such temperature ranges and a nonzero magnetic volume fraction observed by $\mu$SR, may be consequence of the defects that lead to the formation of magnetic polarons.  Future work on this material can be directed towards pressure-dependent experiments, where the spin-lattice interaction can be taken further to explore experimentally the phase diagram for a classical arrangement of spins.  

\section{Conclusion}

We have performed a detailed magnetic characterization on YbMn$_2$Sb$_2$ single crystals, finding a very complex magnetic state, that has been analysed in light of our measurements as well as previous reports and theoretical models.  This work opens a new alternative to study magnetic frustration in general, and will be specially useful to understand and correlate the arrangement of moments in a crystal lattice and their magnetic interactions.  

The authors thank H. Luetkens and A. Amato for their support during the $\mu$SR experiments at PSI, the Multiuser Central Facilities (UFABC) for the experimental support, and E. Morenzoni for important discussions.  The research leading to these results has received funding from the European Community's Seventh Framework Programme (FP7/2007-2013) under Grant Agreement No. 290605 (PSI-FELLOW/COFUND), and from the Brazilian funding agencies CNPq, FAPESP (Grants No. 2011/19924-2 and 2014/20365-6) and CAPES.  RAR was funded by the Gordon and Betty Moore Foundation's EP: QS Initivative through grant No. GBMF4411. 

\nocite{*}

\bibliography{bibliography}

\end{document}